\documentstyle[11pt,epsfig]{article} \oddsidemargin -0.0cm
\begin{document}
\baselineskip 16pt
\title{{\bf Nucleon resonances in $\pi N$ scattering\\ up to energies
$\sqrt{s} \le 2.0$  GeV}}

\author{Guan Yeu Chen$^a$, S.S. Kamalov$^{a,b,c}$, Shin Nan Yang$^a$, D.
Drechsel$^c$, and L. Tiator$^c$\\
$^a$Department of Physics and Center for Theoretical Sciences,\\
National Taiwan University, Taipei 10617, Taiwan\\ $^b$Bogoliubov
Laboratory for Theoretical Physics, JINR, \\Dubna, 141980 Moscow
Region, Russia\\$^c$Institut f\"ur Kernphyik, Universit\"at Mainz,
D-55099 Mainz, Germany }
\date{\today}
\maketitle

\begin{abstract}

A meson-exchange model for  pion-nucleon scattering was previously
constructed using a three-dimensional reduction scheme  of the
Bethe-Salpeter equation for a model Lagrangian involving $\pi$,
$\eta$, $N$, $\Delta$, $\rho$, and $\sigma$ fields. We thereby
extend our previous work by including the $\eta N$ channel and all
the $\pi N$ resonances with masses $\sim 2$ GeV, up to the $F$
waves. The effects of the $\pi\pi N$ channels are taken into
account by introducing an effective width in the resonance
propagators. The extended model gives an excellent fit to both
$\pi N$ phase shifts and inelasticity parameters in all channels
up to the $F$ waves and for energies below 2 GeV. We present a new
scheme to extract the properties of overlapping resonances. The
predicted values for the resonance masses and widths as well as
resonance pole positions and residues are compared to the listing
of the Particle Data Group.\\

\noindent
{\it PACS}:\, 11.80.Gw, 13.75.Gx, 14.20.Gk, 25.80.Dj \\
{\it KEYWORDS}:\, coupled-channel, pion-nucleon interaction,
baryon resonances

\end{abstract}

\newpage

\section{Introduction}

Pion-nucleon scattering is of interest because of its fundamental nature. In
the 1950's, it was widely regarded as {\it the} dynamical problem because of
the special role pions and nucleons play in the family of particles
\cite{gasio66}. Soon it became one of the main sources of information about the
baryon spectrum. The pion-nucleon interaction also plays a fundamental role in
the description of nuclear dynamics for which the $\pi N$ off-shell amplitude
serves as the basic input to most of the existing nuclear calculations at
intermediate energies. Knowledge about the off-shell $\pi N$ amplitude is also
essential in interpreting the experiments performed at the intermediate-energy
electron accelerators in order to unravel the internal structure of these
hadrons. As an example, the importance of the $\pi N$ off-shell $t$-matrix in a
dynamical description of pion electromagnetic production has been demonstrated
in recent years \cite{yang85,lee91,sato96,KY99}. For further progress it is now
necessary to improve our previous description of the $\pi N$ interaction and
to extend it to higher energies. \\

It is commonly accepted that Quantum Chromodynamics (QCD) is the
fundamental theory of the strong interaction. However, due to the
confinement problem, it is still practically impossible  to derive
the $\pi N$ interaction directly from QCD. On the other hand, models
based on meson-exchange pictures \cite{paris,bonn} have been very
successful in describing the $NN$ scattering. Over the last decade,
similarly successful models have also been constructed for $\pi N$
scattering
\cite{lee91,pj,gross93,hung94,juelich,lahiff99,pt00,hung01,gasparyan03,polinder05}.
Most of the recent attempts in this direction were obtained by
applying various three-dimensional reductions of the Bethe-Salpeter
equation, except for Ref.~\cite{lahiff99} in which the
four-dimensional Bethe-Salpeter equation was solved. Because the
effective Lagrangian used in these models includes only the first
few low-lying resonances, in addition to pion and nucleon as well as
$\sigma$ and $\rho$ mesons, the energy region is restricted to low
and
intermediate energies.\\

In previous works we constructed several meson-exchange $\pi N$
models within the Bethe-Salpeter formulation
\cite{lee91,hung94,hung01} and investigated their sensitivity with
respect to various three-dimensional reduction schemes. The model
Lagrangian included only $\pi, N, \Delta, \rho,$ and $\sigma$
fields, and it was found that all the resulting meson-exchange
models can yield similarly good descriptions of the $\pi N$
scattering data up to $400$ MeV. The model obtained with the
Cooper-Jennings reduction scheme \cite{CJ} was recently extended up
to a c.m. energy of 2 GeV in the $S_{11}$ channel by including the
$\eta N$ channel and a set of higher $S_{11}$ resonances
\cite{chen03}. An excellent fit to the t-matrix in both $\pi N$ and
$\eta N$ channels was obtained. In addition, when analyzing the pion
photoproduction data, we obtained background contributions to the
imaginary part of the $S$-wave multipole which differ considerably
from the result based on the $K$-matrix approximation. The resulting
resonance contributions required to explain the pion photoproduction
data led to a substantial change of the extracted electromagnetic
helicity amplitudes. In the present paper we further extend our
model to include the higher partial waves up to the $F$ waves. The
spin-$\frac 32$ resonances are treated as Rarita-Schwinger particles
while we use simple Breit-Wigner forms for the resonance propagators
with spins $\frac 52$ and $\frac 72$. Since the importance of  $\pi
\pi N$ final states grows with energy, also these channels have to
be taken care of. Instead of including them like the $\sigma N$,
$\rho N$, and $\pi\Delta$ states directly in the coupled-channels
calculation as done in Ref.~\cite{gasparyan03}, we follow the recipe
of Ref.~\cite{chen03} to account for the  $\pi\pi N$ channels by
introducing a phenomenological term in the resonance propagators. It
turns out that this approximation works quite well in most of the
considered channels.\\

The question of whether a resonance be a three-quark state dressed
by the meson cloud or be generated dynamically is an issue still
under investigation in the literature. At one extreme, there is the
conjecture \cite{lutz02} that baryon resonances not belonging to the
large-$N_c$ ground states may be generated by coupled-channel
dynamics. On the other hand, in J\"ulich $\pi N$ model
\cite{schutz98,krehl00}, it was found that only the Roper resonance
$P_{11}(1440)$ can be understood in this way, while other resonances
like $S_{11}(1535), S_{11}(1650),$ and $D_{13}(1520)$ had to be
included in the model explicitly, in direct contrast to results of
 \cite{gross93} where Roper resonance is included explicitly but
$S_{11}(1535)$ generated dynamically. Here we take another extreme
and assume all the nucleon resonances are fundamentally three-quark
states dressed by coupling to meson-nucleon channels. Such a picture
has been found to describe well the $\Delta(1232)$
\cite{KY99,kamalov01,pascalutsa07} and $S_{11}$ resonances up to
$2\, GeV$ \cite{chen03} in $\pi N$ scattering and pion
electromagnetic production.\\

In Sec.~II, we summarize the meson-exchange $\pi N$ model
constructed in our previous work. We extend the model to include the
$\eta N$ channel and the higher resonances in Sec.~III. Our results
are presented in Sec.~IV, and some conclusions are given in Sec.~V.

\section{Meson-exchange $\pi N$ model}
\label{sec2}

Let us first outline the content of our previous meson-exchange model
describing the $\pi N$ interaction at low and intermediate
energies~\cite{hung01}. The reaction of interest is
\begin{equation}\label{reaction}
\pi (q) + N(p) \rightarrow \pi (q') + N(p'),
\end{equation}
where $q, p, q^\prime$, and $p^\prime$ are the four-momenta of the respective
particles. We further define the total and relative four-momentum, $P = p+q$
and $k  = p \,\eta_\pi(s)-q \,\eta_N(s)$, respectively, where $s = P^2 = W^2$
is the Mandelstam variable. The dimensionless variables $\eta_\pi(s)$ and
$\eta_N(s)$ represent the freedom in choosing a three-dimensional reduction,
and are constrained by the condition $\eta_N + \eta_\pi = 1$. An often used
definition for these variables is given by
\begin{equation}\label{kinematics}
\eta_N(s)  =  \frac{\varepsilon_N(s)}{\varepsilon_N(s)+\varepsilon_\pi(s)},
\hspace{1.0cm} \eta_\pi(s)  =
\frac{\varepsilon_\pi(s)}{\varepsilon_N(s)+\varepsilon_\pi (s)},
\end{equation}
with $\varepsilon_N(s)  = (s+m_N^2-m_{\pi}^2)/2\sqrt{s}$ and
$\varepsilon_{\pi}(s)  = (s-m_N^2+m_{\pi}^2)/2\sqrt{s}$. For
further details
we refer the reader to Ref.~\cite{hung01}.\\

The Bethe-Salpeter (BS) equation for $\pi N$ scattering takes the general form
\begin{equation}\label{BSeq}
T_{\pi N} = B_{\pi N} + B_{\pi N}G_0 T_{\pi N},
\end{equation}
where $B_{\pi N}$ is the sum of all irreducible two-particle
Feynman amplitudes and $G_0$ the free relativistic pion-nucleon
propagator. The BS equation can be cast into the form
\begin{equation}\label{BSeqa}
 T_{\pi N} = \hat B_{\pi N} + \hat B_{\pi N}\hat G_0T_{\pi N},
\end{equation}
with
\begin{equation}\label{BSeqb}
\hat B_{\pi N} = B_{\pi N} + B_{\pi N}(G_0-\hat G_0)\hat B_{\pi N},
\end{equation}\\
where a three-dimensional reduction of Eq.~(\ref{BSeq}) is obtained by use of
an appropriate propagator $\hat G_0(k;P)$. It is also convenient to choose
$\hat G_0$ such that two-body unitarity is maintained by reproducing the $\pi
N$ elastic cut. There is still a wide range of possible propagators which
satisfy this requirement. A standard choice of the propagator has the
form~\cite{CJ,BS}
\begin{eqnarray}\label{propagatorB}
\hat G_0(k;P) & = & \frac{1}{(2\pi)^3}\int
\frac{ds'}{ s-s'} f(s,s') [\alpha (s,s') P \hspace{-0.10in} / +k
\hspace{-0.08in} / +m_N]  \nonumber  \\
 & & \times \delta^{(+)}([\eta_N(s')P' + k]^2 - m_{N}^2)  \delta^{(+)}
([\eta_\pi (s')P'- k]^2 -  m_{\pi}^2), \label{2GH}
\end{eqnarray}
with $P' = \sqrt{\frac{s'}{s}}P$. The superscript (+) associated with
$\delta$-functions signifies that only the positive energy part is kept in the
propagator. Concerning the Dirac matrices and the Lorentz metrics we use the
notation of Bjorken and Drell~\cite{BD}. Furthermore, the variables $f$ and
$\alpha$ are dimensionless variables containing the freedom of reduction, they
are constrained by the conditions  $ f(s,s)  = 1$ and $\alpha(s,s) =
\eta_N(s)$, which ensure the reproduction of the elastic cut. In the
Cooper-Jennings reduction scheme~\cite{CJ} they take the form

\begin{equation}\label{CJ}
 \alpha(s,s')  =  \eta_N(s), \hspace{2.0cm} f(s,s')=
\frac{4\sqrt{ss'}\varepsilon_N(s')\varepsilon_\pi(s')}{ss'-(m_N^2-m_\pi^2)^2}.
\end{equation}

The integral over $s'$ in Eq.~(\ref{propagatorB}) can be performed. Expressed
in the c.m. frame, the result is
\begin{equation}
\hat G_0(k;s)=\frac{1}{(2\pi)^3} \frac{\delta(k_0-\hat{\eta}(s_{\vec{k}},
\vec{k}))} {\sqrt{s}-\sqrt{s_{\vec{k}}}} \frac{2 \sqrt{s_{\vec{k}}}} {\sqrt{s}+
\sqrt{s_{\vec{k}}}} f(s,s_{\vec{k}}) \frac{\alpha(s,s_{\vec{k}}) \gamma_0
\sqrt{s} +k \hspace{-0.08in} / + m_N} {4 E_N(\vec{k}) E_\pi(\vec{k})},
\end{equation}
where  $E_N(\vec{k})$ and  $E_\pi(\vec{k})$ are the nucleon and
pion energies for the three-momentum $\vec k$, $\sqrt{ s_{\vec{k}}
}  = E_N(\vec{k})+E_\pi(\vec{k}) = E$ is the total energy in the
c.m. frame, and $\hat{\eta}(s,\vec{k}) =
E_N(\vec{k})-\eta_N(s_{\vec{k}}) \sqrt{s_{\vec{k}}}.$ By use of
these relations we obtain the following $\pi N$ scattering
equation:

\begin{equation}\label{3dim-t}
t(\vec{k'},\vec{k};E) = v(\vec{k'},\vec{k};E) + \int d
\vec{k''}v(\vec{k'},\vec{k''};E) g_0(\vec k'';E) t(\vec{k''},\vec{k};E)\, .
\end{equation}

The explicit relations between the variables of Eqs.~(\ref{3dim-t}) and
(\ref{BSeq}) are

\begin{eqnarray}\label{LSeq}
t(\vec{k'},\vec{k};E) & = &\int dk'_0 dk_0\delta(k'_0-\hat{\eta}') T(k',k;E)
\delta(k_0-\hat{\eta}), \nonumber  \\
v(\vec{k'},\vec{k};E) & = &\int dk'_0 dk_0\delta(k'_0-\hat{\eta}') B(k',k;E)
\delta(k_0-\hat{\eta}), \nonumber \\
g_0(\vec k;E) & = & \int dk_0 \hat G_0(k;E),
\end{eqnarray}

with $\hat{\eta}'=  \hat{\eta}(s_{\vec{k'}}, \vec{k'})$ and $\hat{\eta}=
\hat{\eta}(s_{\vec{k}}, \vec{k})$.\\

Because our previous work concentrated on the  $\pi N$ scattering process at
low and intermediate energies, we only considered the degrees of freedom due to
the $\pi, N, \sigma, \rho,$ and $\Delta(1232)$ fields, and approximated the sum
of all irreducible two-particle Feynman amplitudes, $B(k',k;E)$ in
Eq.~(\ref{LSeq}), by the tree approximation of the following interaction
Lagrangian:
\begin{eqnarray}\label{Lagrangian}
{\cal L}_I & = & \frac{f_{\pi NN}^{(0)}}{m_\pi} \bar N \gamma_5 \gamma_\mu
\vec{\tau} \cdot \partial^\mu \vec{\pi} N
 - g^{(s)}_{\sigma\pi\pi} m_{\pi} \sigma (\vec{\pi}\cdot\vec{\pi})
 - \frac {g^{(v)}_{\sigma\pi\pi}}{2  m_{\pi}} \sigma
 (\partial^{\mu}\vec{\pi}\cdot\partial_{\mu}\vec{\pi})
 -g_{\sigma NN}\bar{N}\sigma N \nonumber \\
& & -g_{\rho NN} \bar{N} \{ \gamma_\mu \vec{\rho}\,{}^\mu +
\frac{\kappa_V^\rho}{4m_N} \sigma_{\mu\nu} (\partial ^\mu \vec{\rho}\,{}^\nu -
\partial^\nu \vec{\rho}\,{}^\mu) \} \cdot \frac{1}{2}\vec{\tau} N  \nonumber \\
& & -g_{\rho\pi\pi} \vec{\rho}\,{}^{\mu} \cdot (\vec{\pi} \times \partial_{\mu}
\vec{\pi}) - \frac{g_{\rho\pi\pi}}{4m_{\rho}^2}(\delta - 1)(\partial^\mu
\vec{\rho}\,{}^\nu -\partial^\nu \vec{\rho}\,{}^\mu) \cdot (\partial_\mu
\vec{\pi} \times\partial_\nu \vec{\pi})  \nonumber \\
& & + \frac{g_{\pi N\Delta}}{m_\pi}  \bar\Delta_\mu
[g^{\mu\nu} -(Z+\frac{1}{2})\gamma^\mu\gamma^\nu] \vec{T}_{\Delta N}N \cdot
\partial_{\nu} \vec \pi,
\end{eqnarray}
with  $\Delta_{\mu}$ the Rarita-Schwinger field operator for the $\Delta$
resonance and $\vec T_{\Delta N}$ the isospin transition operator between the
nucleon and the $\Delta$. The resulting driving term consists of the direct and
crossed $N$ and $\Delta$ diagrams as well as the t-channel $\sigma$- and
$\rho$-exchange contributions. \\

The procedure of Afnan and collaborators~\cite{Morioka} was followed to
constrain the $P_{11}$ phase shift by imposing the nucleon pole condition. This
treatment leads to a proper renormalization of both nucleon mass and $\pi NN$
coupling constant. It also yields the important cancelation between the
repulsive nucleon pole contribution and the attractive background, such that a
reasonable fit to the $\pi N$ phase shifts in the $P_{11}$ channel can be achieved.\\

To complete the model we further introduced form factors to
regularize the driving term $v(\vec{k},\vec{k}^\prime)$ of
Eq.~(\ref{LSeq}). For this purpose covariant form factors of the
form
\begin{equation}\label{ff}
F(p^2)=[\frac {n\Lambda^4}{n\Lambda^4 + (m^2 -
p^2)^2}]^{n},
\end{equation}
are associated with each legs of the vertices, where $p$ is the
four-momentum and $m$ the mass of the respective particle. This
parameterization is similar to the prescription of Ref.~\cite{pj}
and in Ref.~\cite{hung01} both $n = 10$ and $2$ were considered.
However, in our
previous work we used the value $n = 10$ \cite{chen03}.\\

The parameters that were allowed to vary in fitting the empirical phase shifts
are the products $g_{\sigma NN}g^{(s)}_{\sigma\pi\pi}, \, g_{\sigma
NN}g^{(v)}_{\sigma\pi\pi}$, and $g_{\rho NN}g_{\rho\pi\pi}$ as well as $\delta$
for the t-channel $\sigma$ and $\rho$ exchanges, $m_{\Delta}^{(0)},\,
g^{(0)}_{\pi N\Delta}$, and $Z$ for the $\Delta$ mechanism, and the cut-off
parameters $\Lambda$ of the form factors given by Eq.~(\ref{ff}). In the
crossed $N$ diagram, the physical $\pi NN$ coupling constant is used. For the
crossed $\Delta$ diagram, the situation is not so clear since the determination
of the "physical" $\pi N\Delta$ coupling constant depends on the non-resonant
contribution in the $P_{33}$ channel. In principle, it can be determined by
carrying out a renormalization procedure similar to that used for the nucleon.
However, this would require a much more difficult numerical task, because the
$\Delta$ pole is complex. In accordance with Refs.~\cite{lee91,pj,gross93}, we
therefore did not carry out such a renormalization for the $\Delta$ but simply
determined the coupling constant in the crossed $\Delta$ diagram by
a fit to the data. The resulting coupling constant was denoted as $g_{\pi N\Delta}$.\\

\section{Extension to higher energies: inclusion of the $\eta N$ channel and higher resonances}
\label{sec3}

As the energy increases, two-pion channels like $\sigma N, \eta N, \pi\Delta,
\rho N$ as well as a non-resonant continuum of $\pi \pi N$ states become
increasingly important, and at the same time more and more nucleon resonances
appear as intermediate states. The $\pi N$ model described in Sec.~\ref{sec2}
was therefore extended for the $S_{11}$ partial wave by explicitly coupling the
$\pi$, $\eta$ and $\pi \pi$ channels and including the couplings with higher
baryon resonances~\cite{chen03}. In particular, in the case of only one
contributing resonance $R$, the Hilbert space was enlarged by the inclusion of
a bare $S_{11}$ resonance $R$ which acquires a width by its coupling with the
$\pi N$ and $\eta N$ channels through the Lagrangian
\begin{eqnarray}
{\mathcal{L_I}}= ig^{(0)}_{\pi NR}\bar R \tau N\cdot \pi +
ig^{(0)}_{\eta NR}\bar R N\eta + h.c., \label{lagr}
\end{eqnarray}
where $N, R, \pi,$ and $\eta$ denote the field operators for the nucleon, bare
resonance $R$, pion, and eta meson, respectively. The full $t$-matrix can be
written as a system of coupled equations,
\begin{eqnarray}
t_{ij}(E)= v_{ij}(E)+\sum_k  v_{ik}(E)\,g_k(E)\, t_{kj}(E)\,,
\label{t_ij}
\end{eqnarray}
with $i$ and $j$ denoting the $\pi$ and $\eta$ channels and $E=W$ is
the total c.m. energy.\\

In general, the potential $ v_{ij}$ is the sum of non-resonant
$(v^B_{ij})$ and bare resonance $(v^R_{ij})$ terms,
\begin{eqnarray}
v_{ij}(E)=  v^B_{ij}(E)+ v^R_{ij}(E)\,. \label{v_ij}
\end{eqnarray}
The non-resonant term $v^B_{\pi\pi}$ for the $\pi N$ elastic
channel is given by the results of  Sec.~\ref{sec2} and contains
contributions from the $s$- and $u$-channels, Born terms and
$t$-channel contributions with $\omega$, $\rho$, and $\sigma$
exchange. The parameters in $v^B_{\pi\pi}$ are fixed from the
analysis of the pion scattering phase shifts for the $s-$ and
$p-$waves at low energies ($W<1300$ MeV) \cite{hung01}. In
channels involving the $\eta$, the potential $v^B_{i\eta}$ is
taken to be zero because the $\eta NN$ coupling is very small
\cite{TBK}.

The bare resonance contribution arises from the excitation and de-excitation of
the resonance $R$,
\begin{eqnarray} v^R_{ij}(E)=\frac
{h_{i R}^{(0)\dagger} h_{j R}^{(0)}}{E-M_R^{(0)}}, \label{vRbare}
\end{eqnarray}
where $h_{i R}^{(0)}$ and $M_R^{(0)}$ denote the bare vertex operator for
$R\rightarrow \pi/\eta + N$ and the bare mass of the resonance $R$,
respectively. The matrix elements of the potential $v^R_{ij}(E)$ can be
symbolically expressed in the form
\begin{eqnarray}
v^R_{ij}(q,q';E)=\frac{f_i(\tilde
{\Lambda}_i,q;E)\,g_i^{(0)}\,g_j^{(0)}\,
f_j(\tilde{\Lambda}_j,q';E)}{E-M_R^{(0)}+
\frac{i}{2}\Gamma_R^{2\pi}(E)} \,, \label{v_R1}
\end{eqnarray}
where $q$ and $q'$ are the pion (or eta) momenta in the initial
and final states, and $g_{i/j}^{(0)}$ is the resonance vertex
couplings. As in \cite{hung01}, we associate with each external
line of the particle $\alpha$ in a $\pi NR$ vertex a covariant
form factor $F_\alpha =
[n\Lambda^4_\alpha/(n\Lambda^4_\alpha+(p^2_\alpha
-m^2_\alpha)^2)]^{n}$, where $p_\alpha$, $m_\alpha$, and
$\Lambda_\alpha$ are the four-momentum, mass, and cut-off
parameter of particle $\alpha$, respectively, and $n=10$. As a
result, $f_i$ depends on the  product of three cut-off parameters,
i.e.,
$\tilde\Lambda_\pi \equiv (\Lambda_N,\Lambda_R,\Lambda_\pi)$.\\

In Eq.~(\ref{v_R1}) we have included a phenomenological term
$\Gamma_R^{2\pi}(E)$ in the resonance propagator to account for the $\pi\pi N$
decay channel. Therefore, our "bare" resonance propagator already contains some
renormalization or "dressing" effects due to the coupling with the $\pi\pi N$
channel. With this prescription we assume that any further non-resonant
coupling mechanism with the $\pi\pi N$ channel is small. Following
Refs.~\cite{Lvov,MAID98} we take
\begin{eqnarray}
\Gamma_R^{2\pi}(E)=\Gamma^{2\pi(0)}_{R}\left(\frac{q_{2\pi}}{q_0}\right)^{2l+4}
\left(\frac{X_R^2+q^2_0}{X_R^2+q^2_{2\pi}}\right)^{l+2}\,,
\label{G_2pi}
\end{eqnarray}
where $l$ is the pion orbital momentum, $q_{2\pi}=q_{2\pi}(E)$ the
momentum of the compound two-pion system,
$q_0=q_{2\pi}(E=M_{R}^{(0)})$ and the quantity
$\Gamma^{2\pi(0)}_{R}$ is the $2\pi$ decay width at resonance. We
note that this form accounts for the correct energy behavior of
the phase space near the three-body threshold~\cite{Lvov}. In our
present work, $\Gamma_{R}^{2\pi(0)}$ and $X_R$ are considered as a
free parameters. As a result, one isolated resonance will in
general contain six free parameters, the bare mass $M^{(0)}_R$,
the decay width $\Gamma^{2\pi(0)}_{R}$, two bare coupling
constants $g^{(0)}_i$ and $g^{(0)}_j$, and two cut-off parameters
$\Lambda_R$ and $X_R$. The generalization of the coupled channels
model to the case of $N$ resonances with the same quantum numbers
is then given by
\begin{eqnarray}
v^R_{ij}(q,q';E)=\sum_{n=1}^{N} v^{R_n}_{ij}(q,q';E), \label{v_RN}
\end{eqnarray}
with free parameters for the bare masses, widths, coupling
constants, and
cut-off parameters for each resonance.\\

Having solved the coupled channel equations, our next task is the extraction of
the physical (or "dressed") masses, partial widths, and branching ratios of the
resonances. It is known that this procedure is model dependent because the
background and the resonance contributions can not be separated in a unique
way. Of course, the solution to this problem becomes more and more difficult
with an increasing number of overlapping resonances in the same channel. In the
literature, there are two schemes used to separate the total $t-$matrix into
background and resonance contributions. For simplicity, we illustrate these two
methods for the uncoupled channel case, i.e., assuming that only the $\pi N$
channel is open. In this case the potential operator describing the excitation
of a bare resonance $R$ takes the form
\begin{eqnarray}
v^{R}_{\pi N}(E)=\frac {h_{\pi R}^{(0)\dagger} h_{\pi
R}^{(0)}}{E-M_R^{(0)}}, \label{vRbare}
\end{eqnarray}
with $h_{\pi R}^{(0)}$ the bare vertex.\\

The first scheme was suggested by Afnan and
collaborators~\cite{morioka} and recently used in the dynamical
model calculation of pion scattering and pion photoproduction
\cite{sato96}. By use of the two-potential formulation the
$t-$matrix is written as
\begin{eqnarray}
t_{\pi N}(E)=\tilde{t}_{\pi N}^B(E) + \tilde{t}_{\pi N}^R(E),
\label{tsl}
\end{eqnarray}
where $\tilde{t}_{\pi N}^B(E)$ is defined as

\begin{eqnarray}
\tilde{t}_{\pi N}^B(E)=v_{\pi N}^B+v_{\pi N}^B\,g_0(E) \,\tilde{t}_{\pi
N}^B(E). \label{tilde-tB}
\end{eqnarray}

\begin{figure}[h]
\begin{center}
\epsfig{file=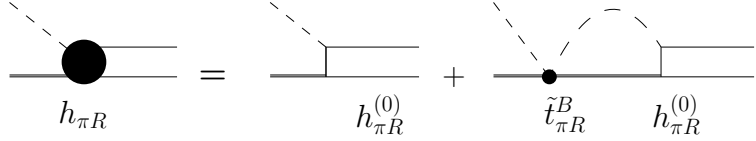,width=10cm}
\end{center}
\caption{Dressed and bare $\pi N R$ vertex}
\label{fig1-vertex} \vspace{-0.5cm}
\end{figure}

\begin{figure}[h]
\begin{center}
\epsfig{file=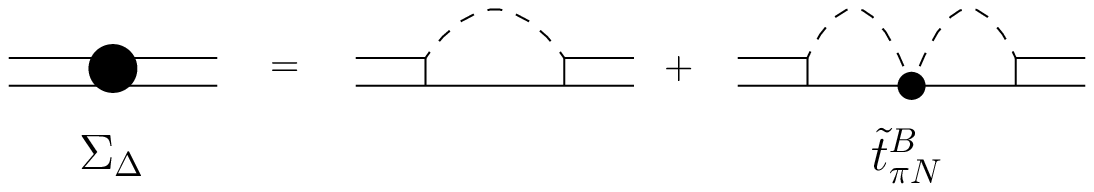,width=10cm}
\end{center}
\caption{Resonance self-energy } \label{fig2-selfenergy}
\end{figure}

We will call $\tilde{t}_{\pi N}^B(E)$ the "non-resonant" background because it
does not contain any resonance contribution from $v_{\pi N}^{R}$ of
Eq.~(\ref{vRbare}). The resonance term $\tilde{t}_{\pi N}^R(E)$ takes the form
\begin{eqnarray}
\tilde{t}_{\pi N}^R(E)= \bar h_{\pi R}(E)
\frac{1}{E-M_{R}^{(0)}-\Sigma_{R}(E)} h_{\pi R}(E),\label{tilde-tR}
\end{eqnarray}
with the definitions

\begin{eqnarray}
h_{\pi R}(E)=h_{\pi R}^{(0)}+h_{\pi R}^{(0)}\,g_0(E)
\,\tilde{t}_{\pi N}^{B}(E) \, , \\
\label{hpiN_Rpisl}\bar h_{\pi R}(E)=h_{\pi
R}^{(0)\dagger}+\tilde{t}_{\pi N}^B(E)\,g_0(E) \,h_{\pi
R}^{(0)\dagger} \, , \label{hpiN_Rpisl1}
\end{eqnarray}
and the self-energy $\Sigma_{R}(E)$ given by
\begin{eqnarray}
\Sigma_{R}(E)=h_{\pi R}^{(0)}\, g_0 \,\bar h_{\pi R}(E)= h_{\pi
R}^{(0)}\,g_0\, h_{\pi R}^{(0)\dagger}+ h_{\pi R}^{(0)}\,g_0\,
\tilde{t}_{\pi N}^B(E)\, g_0 \,h_{\pi R}^{(0)\dagger} \,.
\label{Sigma_Delta}
\end{eqnarray}
Graphical representations of the dressed vertex $h_{\pi R}(E)$ and
the self-energy $\Sigma_{R}(E)$ are depicted in
Figs.~\ref{fig1-vertex} and \ref{fig2-selfenergy}, respectively,
where the solid circle on the l.h.s. of Fig.~\ref{fig1-vertex}
denotes the dressed vertex $h_{\pi R}$ while the $\pi NR$ vertices
, $h_{\pi R}^{(0)}$, on the r.h.s. of Fig.~\ref{fig1-vertex}
correspond to the excitation of a bare resonance $R$. The small
solid circles in Figs.~\ref{fig1-vertex} and \ref{fig2-selfenergy}
represent the "non-resonant" background $\tilde{t}_{\pi N}^B(E)$
as defined in Eq.~\ref{tilde-tB}.

The information about the physical mass and the total width of the resonance
$R$ are contained in the dressed resonance propagator given in
Eq.~(\ref{tilde-tR}). The complex self-energy $\Sigma(E)$ leads to a shift from
the real "bare" mass to a complex and energy-dependent value. However, we
characterize the resonances by energy-independent parameters that are obtained
by solving the equation

\begin{eqnarray}
E-M_{R}^{(0)}-Re \, \Sigma_{R}(E)=0  \, . \label{Delta_mass1}
\end{eqnarray}.

The solution of this equation, $E=M_R$, corresponds to the energy at
which the dressed propagator in Eq.~(\ref{tilde-tR}) becomes purely
imaginary and is used to  define the "physical" or "dressed" mass,

\begin{eqnarray}\label{Mdressed}
 M_{R}=M_{R}^{(0)}+Re\,\Sigma_{R}(M_{R}) \, \label{dressed_mass},
\end{eqnarray}
and the width of the resonance

\begin{eqnarray}\label{width}
\Gamma_{R}(M_R)=-2\,Im\,\Sigma_{R}(M_R) \,\label{width}.
\end{eqnarray}

We hasten to add that in some of the literature the resonance mass is defined
by the energy at which the phase of the resonance contribution
$\tilde{t}^{R}_{\pi N}$ passes through $90^{\circ}$. Because of the
non-resonant background, however, the numerator of $\tilde{t}^{R}_{\pi N}$
becomes a complex number with the phase $2\delta^B$. Therefore, the position of
the resonance and consequently the total width will now differ from our
Eqs.~(\ref{Mdressed}) and (\ref{width}). We believe that the physical masses
and widths defined in Eqs.~(\ref{Delta_mass1})-(\ref{width})
are more closely related to what one would eventually calculate in lattice QCD.\\

All of the above equations are based on the two-potential
formulation~\cite{Watson}. The extension of this method to the case
of several overlapping resonances in the same partial channel
$\alpha$ complicates the problem. In particular, we can not express
the $t-$matrix as a simple sum of a smooth background and
overlapping resonances,
\begin{eqnarray}
t_{\pi N}(E)\neq \tilde{t}^{B}_{\pi N}(E)+ \sum_{i=1}^{N}
\tilde{t}^{R_i}_{\pi N}(E)\,. \label{T_piN}
\end{eqnarray}

We therefore prefer to separate the resonance and background
contributions in the framework of Refs.~\cite{KY99,chen03}. In
this approach the full pion-nucleon scattering matrix is
decomposed as follows:
\begin{eqnarray}
t_{\pi N}(E)=t_{\pi N}^B(E) + t_{\pi
N}^{R}(E),\label{eq:tgammapi33}
\end{eqnarray}
where
\begin{eqnarray}
t_{\pi N}^B(E)&=&v_{\pi N}^B+v_{\pi N}^B\,g_0(E)\,t_{\pi N}(E)\,,
\label{DMT_bcr}\\
t_{\pi N}^{R}(E)&=&v_{\pi N}^R+v_{\pi N}^R\,g_0(E) \,t_{\pi N}(E).
\label{DMT_res}
\end{eqnarray}

Comparing  $t_{\pi N}^B$ with  ${\tilde t}_{\gamma\pi}^B$ of
Eq.~(\ref{tilde-tB}), the "background" $t_{\pi N}^B$ now includes
contributions not only from the background rescattering but also
from intermediate resonance excitation. This is compensated by the
fact that the resonance contribution $t_{\pi N}^{R}$ now contains
only the terms that start with the bare resonance excitation.
Expressed in terms of self-energy and vertex functions, we obtain
the result
\begin{eqnarray}
t_{\pi N}^{R}(E)=\frac{\bar h_{\pi R}(E) h_{\pi
R}^{(0)}}{E-M_{R}^{(0)}(E) -\Sigma_{R}(E)} \, , \label{tR_DMT2}
\end{eqnarray}
which differs from Eq.~(\ref{tilde-tR}) where dressed vertex appears
in both the initial and final states. On the other hand, we note
that the resonance propagators of the two approaches are identical.
Therefore, the physical masses and total widths determined in the
two methods will be the same.

The second method can be easily extended to the case of $N$
overlapping resonances by the following decomposition of the full
$\pi N$ scattering matrix into background and resonance
contributions,
\begin{eqnarray}
t_{\pi N}(E)= t^{B}_{\pi N}(E)+ \sum_{i=1}^{N}  t^{R_i}_{\pi N}(E)\,.
\label{T_piN}
\end{eqnarray}
The contribution from each resonance $R_i$ can be expressed in terms
of the bare $h_{\pi R_i}^{(0)}$ and dressed $h_{\pi R_i}(E)$ vertex
operators as well as the resonance self energy derived from one-pion
$\Sigma^{1\pi}_{R_i}(E)$ and two-pion $\Sigma^{2\pi}_{R_i}(E)$
channels, that is
\begin{eqnarray}
t_{\pi N}^{R_i}(E)=\frac{\bar h_{\pi R_i}(E) h_{\pi R_i
}^{(0)}}{E-M_{R_i}^{(0)}-\Sigma^{1\pi}_{R_i}(E)-\Sigma^{2\pi}_{R_i}(E)}\,,
\label{tRidressed}
\end{eqnarray}
where $M_{R_i}^{(0)}$ is the bare mass of the $i$-th resonance. The
contributions from the two-pion channel, $\Sigma^{2\pi}_{R_i}$, is
defined phenomenologically as in Eq.~(\ref{G_2pi}). The vertices for
the resonance excitation are obtained from the following equations,
\begin{eqnarray}
h_{\pi R_i}(E)&=&h_{\pi R_i}^{(0)}+h_{\pi R_i}^{(0)}\,g_0(E)
\,t_{\pi N}^{B_i}(E)  \label{hRi}\, ,\\
\bar h_{\pi R_i}(E)&=&h_{\pi R_i}^{(0)\dagger}+t_{\pi N}^{B_i}(E)
\,g_0(E)\,h_{\pi R_i}^{(0)\dagger} \, , \label{DMT_bcr}
\end{eqnarray}
where
\begin{eqnarray}
t_{\pi N}^{B_i}(E)&=&v_{i}(E)+v_i(E)\,g_0(E)\,t_{\pi N}^{B_i}(E),\label{t_piNB_i}\\
v_i(E)&=&v_{\pi N}^B+\sum_{j\neq i}^{N}\,v_{\pi
N}^{R_j}(E)\label{v_i},
\end{eqnarray} with
$v_{\pi N}^{R_j}(E)$ arising from the excitation of the resonance
$R_j$ as given in Eq.~(\ref{v_R1}). The one-pion self-energies
arising from $t_{\pi N}^{B_i}(E)$ of Eq.~(\ref{t_piNB_i}) are
given as
\begin{eqnarray}
\Sigma^{1\pi}_{R_i}(E)=h_{\pi R_i}^{(0)}\, g_0 \,\bar h_{\pi R_i}(E)= h_{\pi
R_i}^{(0)}\,g_0\, h_{\pi R_i}^{(0)\dagger}+ h_{\pi R_i}^{(0)}\,g_0\, t_{\pi
N}^{B_i}\, g_0 \,h_{\pi R_i}^{(0)\dagger} \, , \label{Sigma_Delta}
\end{eqnarray}
and the one-pion branching ratio at the dressed resonance is
\begin{eqnarray}\label{beta}
 \beta^{1\pi}_i=\frac{\Sigma^{1\pi}_{R_i}(M_R)}{\Sigma^{1\pi}_{R_i}(M_R)
 +\Sigma^{2\pi}_{R_i}(M_R)} \,\label{beta_i} .
\end{eqnarray}

The pole positions in the complex energy plane and the complex residues of the
scattering amplitudes at these poles are calculated using the speed plot
technique for the pion-nucleon partial waves. For details see
Refs.~\cite{Hoe92,Han96}.

Finally, it is not difficult to see from Eqs.~(\ref{hRi}) and
(\ref{DMT_bcr}) that the matrix elements of both $h_{\pi R_i}$ and
$t_{\pi N}^{B_i}(E)$ would have the same phase $\phi_R(E)$, i.e.,
\begin{eqnarray}
<h_{\pi R_i}(E)>=\mid <h_{\pi R_i}(E)>\mid
\exp{(i\phi_{R_i})},\qquad t_{\pi N}^{B_i}(E)=\mid <t_{\pi
N}^{B_i}(E)> \mid \exp{(i\phi_{R_i})}\,,\label{phi_Ri}
\end{eqnarray}
where the parenthesis $<>$ is used to denote matrix element of the operator
sandwiched between same set of initial and final states. It then follows that
also the numerator on the r.h.s. of Eq.~(\ref{tRidressed}) carries the phase
$\phi_{R_i}(E)$. Information about this phase is very important for the
phenomenological Breit-Wigner
parametrization of the resonance contributions.\\

We emphasize that in the formulation of
Eqs.~(\ref{T_piN}-\ref{Sigma_Delta}), the nucleon resonances are
treated in a completely symmetrical way. In addition, the
self-energy and the dressing of any resonance receive
contributions from all other resonances.

\section{Results and discussion}
\subsection{$\pi N$ scattering amplitudes}
\begin{figure}[htb]
\begin{center}
\epsfig{file=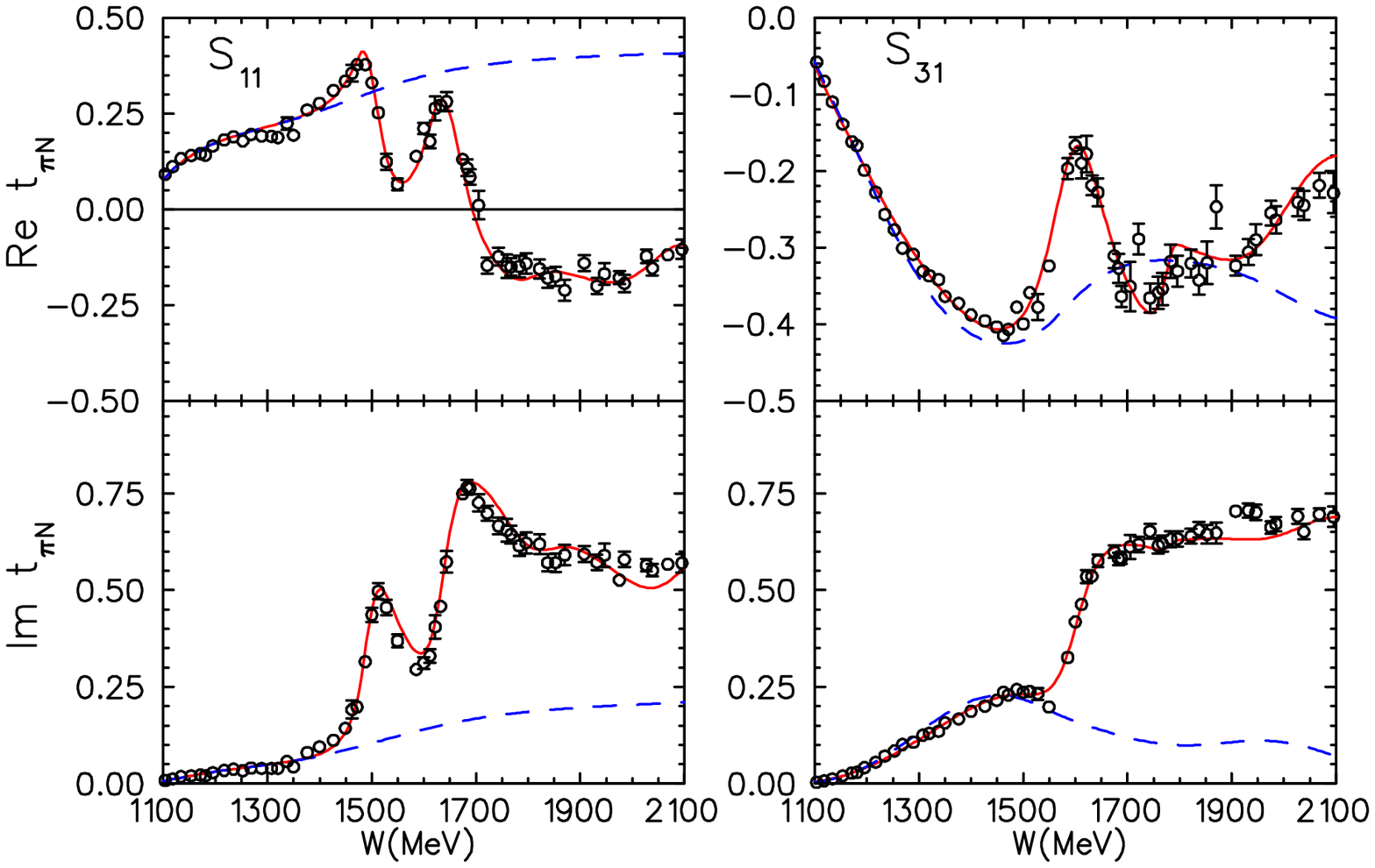, width=12cm}
\end{center}
\caption{The real and imaginary parts of $t_{\pi N}$ in the
S-waves as function of the total c.m. energy $W$. The solid (red)
lines are the best fits of our meson-exchange model, the dashed
(blue) lines correspond to the contributions of non-resonant
background $\tilde t^B_{\pi N}$ of Eq.~(\ref{tilde-tB}). The open
circles are the results of partial waves analysis of
Ref.~\cite{SAID04}.} \label{S-waves}
\end{figure}

\begin{figure}[htb]
\begin{center}
\epsfig{file=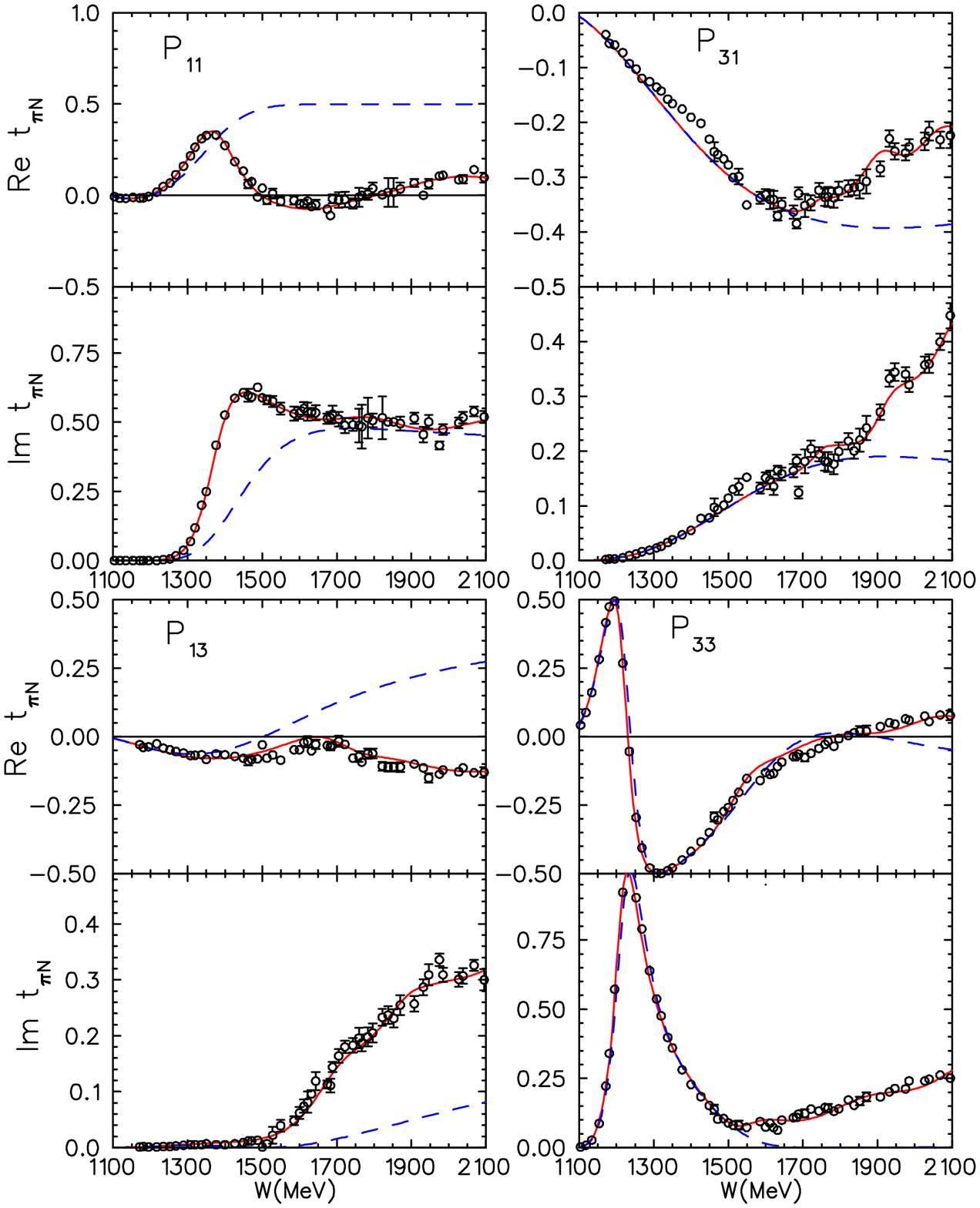, width=12cm}
\end{center}
\caption{The real and imaginary parts of $t_{\pi N}$ in the P-waves as function
of the total c.m. energy $W$. Notation as in Fig.~\ref{S-waves}.}
\label{P-waves} \vspace{-0.5cm}
\end{figure}

\begin{figure}[htb]
\begin{center}
\epsfig{file=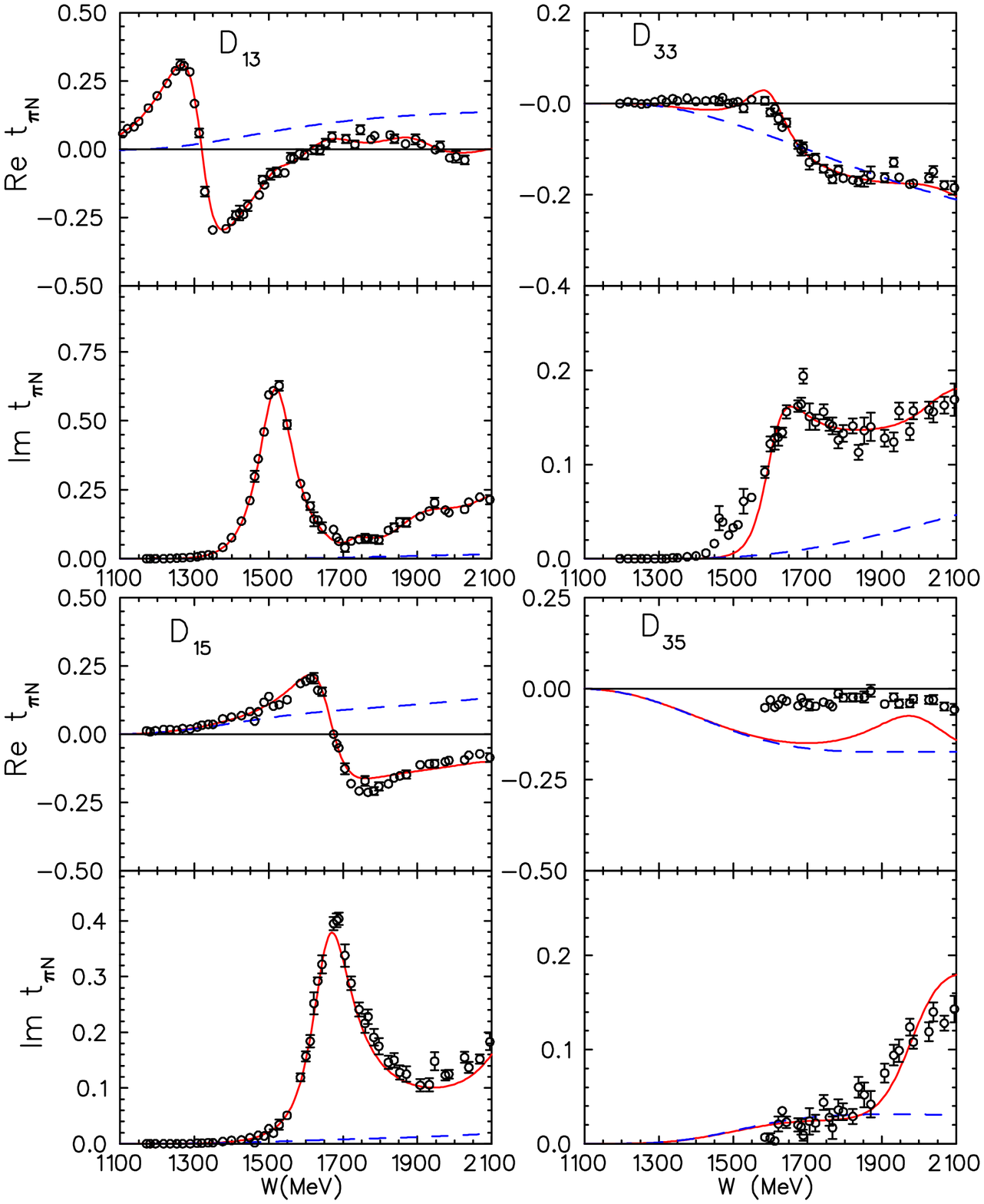, width=12cm}
\end{center}
\caption{The real and imaginary parts of $t_{\pi N}$ in the D-waves as function
of the total c.m. energy $W$. Notation as in Fig.~\ref{S-waves}.}
\label{D-waves} \vspace{-0.5cm}
\end{figure}

\begin{figure}[htb]
\begin{center}
\epsfig{file=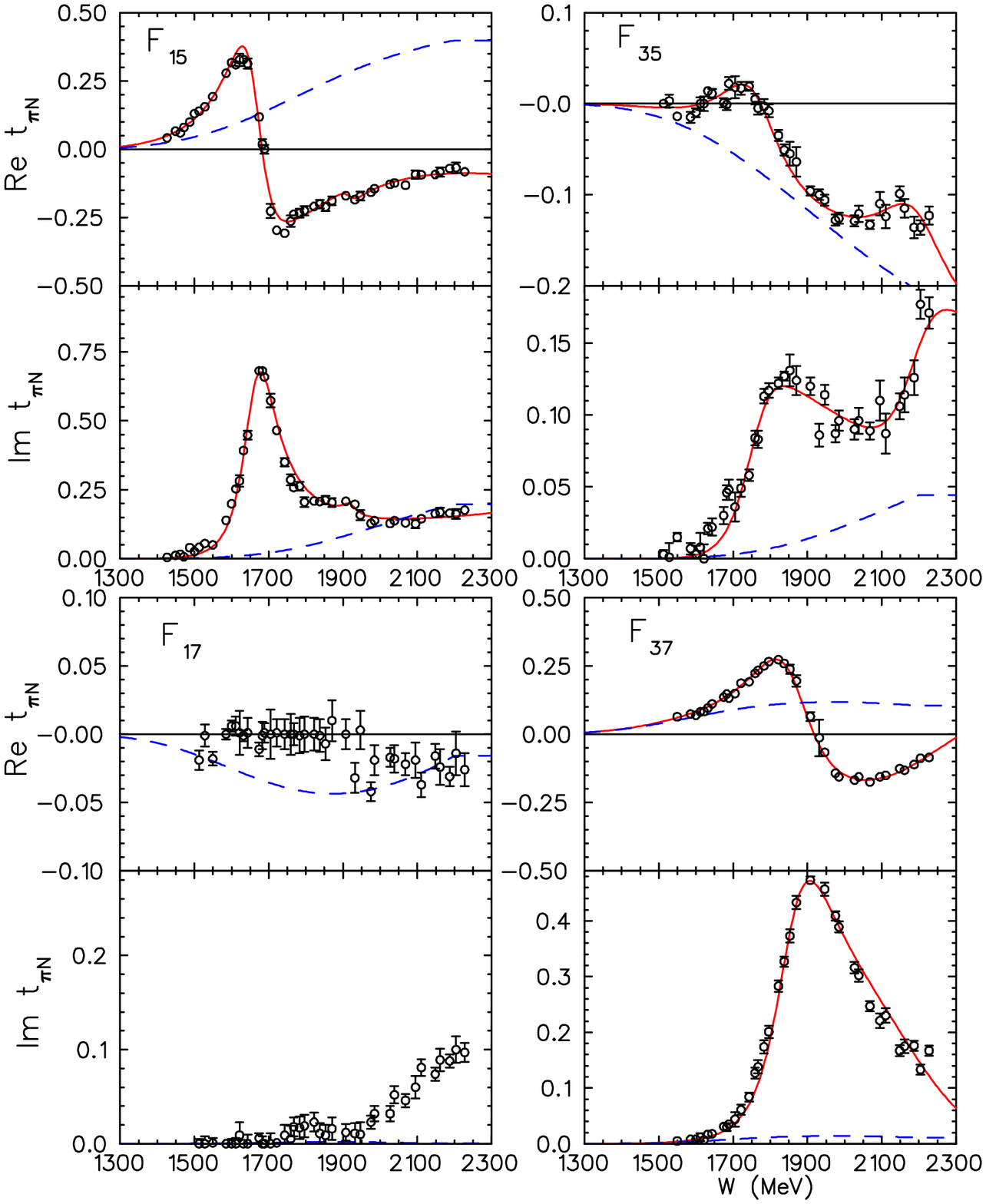, width=12cm}
\end{center}
\caption{The real and imaginary parts of $t_{\pi N}$ in the F-waves as function
of the total c.m. energy $W$. Notation as in Fig.~\ref{S-waves}.}
\label{F-waves} \vspace{-0.5cm}
\end{figure}

With the extended meson exchange model described in
Sec.~\ref{sec3}, we have fitted the $\pi N$ phase shifts and
inelasticity parameters in all channels up to the $F$-waves and
for energies less than 2~GeV. The results for the real and
imaginary parts of the partial wave amplitudes $t_{\pi N}$ are
shown in Figs.~\ref{S-waves}-\ref{F-waves}. The solid lines are
the best fit within our model, while the dashed lines correspond
to the non-resonant background $\tilde t^B_{\pi N}$ of
Eq.~(\ref{tilde-tB}). The open circles represent the data
according to Ref.~\cite{SAID04}. These figures show an excellent
description for both the real and the imaginary parts of the
pion-nucleon scattering amplitudes in all cases except for the
$D_{35}$ and $F_{17}$ channels. For the $D_{35}$, our problem lies
mostly within the real part as seen in Fig.~\ref{D-waves}. In the
case of the $F_{17}$ in Fig. \ref{F-waves}, the inclusion of
further resonances does not improve on the non-resonant background
shown by the data, neither for the real nor for the imaginary part
of the scattering amplitude.\\

\subsection{Resonance parameters}
Let us now look at the resonance parameters whose determination was one of the
main issues of our investigation. Before going into details by comparing with
the PDG values, we point out that our data analysis requires four very broad
resonances, $S_{11}~(1878), D_{13}~(2152), P_{13}~(2204)$, and $P_{31}~(2100)$,
states that are not in the current listing of the PDG~\cite{PDG2006}.
Furthermore, we can not remove the discrepancy between the background
contributions and the data in the $F_{17}$ channel by adding the
$F_{17}(1990)$-resonance listed by the PDG, which is in line with the results
of the SAID analysis~\cite{SAID04}.\\

The physical mass $M_R$, total width $\Gamma_R$, single-pion
branching ratio $\beta_R^{1\pi}$, and background phase ${\phi_R}$
defined for each overlapping nucleon resonance $R$ have been
determined from Eqs.~(\ref{dressed_mass}-\ref{width}) and
(\ref{beta_i}-\ref{phi_Ri}). The results are presented in
Tables~\ref{table:param-1/2} and \ref{table:param-3/2} for the
isospin-$\frac{1}{2}$ and isospin-$\frac{3}{2}$ resonances,
respectively. Using the speed-plot technique, we have also
calculated the resonance pole positions in the complex energy
plane and the complex residues at these poles. The results are
listed in Tables~\ref{table:pole-1/2} and \ref{table:pole-3/2} for
the isospin-$\frac{1}{2}$ and isospin-$\frac{3}{2}$ resonances,
respectively. In Tables \ref{table:param-1/2}-\ref{table:pole-3/2}
we also compare our results
to the listings of the PDG.\\

A word of caution is necessary when comparing the resonance
parameters obtained in our present work with the PDG values. In
many investigations the resonance mass is defined as the energy at
which the phase of the resonance contribution, $\tilde t^R_{\pi
N}$ given by Eq.~(\ref{tilde-tR}), takes the value $\pi/2$.
However, in our present work we define the resonance position as
the energy at which the phase of the denominator in
Eq.~(\ref{tilde-tR}) runs through $90^\circ$, i.e., as the
solution of $(E-M_R^{(0)}-\Sigma_R(E))=0$ given in
Eq.~(\ref{Delta_mass1}). In our opinion this definition has a
better physical interpretation.\\

\noindent{\bf S-waves}\\ As reported in Ref.~\cite{chen03}, we need four
$S_{11}$ resonances to fit the $\pi N$ scattering amplitude in this channel,
instead of the three resonances listed by the PDG. The additional resonance
$S_{11}(1878)$ was found to play an important role in pion photoproduction as
well~\cite{chen03}, but was not seen in both the $\pi N \rightarrow\eta N$
reaction and recent measurements of $\eta$ photoproduction from the
proton~\cite{Thoma05}. There also arise some differences in the resonance
parameters between our present results and those given by Ref.~\cite{chen03}
because of the different definitions for the resonance masses and widths
explained in the previous text. It turns out that the choice of these
definitions has little effect on the extracted masses of all four $S_{11}$
resonances. However, the extracted widths for the first and third resonances
depend very much on the definitions, which leads to an increase of the width
with regard to earlier work, from 90~MeV to 130~MeV and from 265~MeV to
508~MeV, respectively. Our results obtained for the pole position via the speed
plot technique generally agree with the PDG values for the real parts of the
pole positions. However, we obtain much smaller values for the imaginary parts
of the pole positions and also for the residues at the pole.\\

For the isospin-3/2 channel, our extracted masses and widths differ from the
PDG values by more than 100 MeV, all except the first resonance
$S_{31}~(1620)$. The values obtained for the pole positions agree with the PDG
values for the lower resonances. However, the imaginary part of the pole
position for the $S_{31}~(1900)$ and also its residue at the pole comes out very small.
\\

\noindent{\bf P-waves}\\ For P-waves with isospin-1/2, our results are in good
agreement with the PDG values regarding both pole positions and residues.
However, the extracted widths are much larger than the corresponding PDG
values. We also need an extra resonance $P_{13}(2204)$ in order to fit the
scattering amplitude in this channel.\\

For the isospin-3/2 resonance $P_{31}~(1750)$  and $P_{33}(1920)$,
we extract widths of about 500~MeV and 800~MeV, respectively, both
very much above the PDG values. On the other hand, the residue and
the imaginary part of the pole position for $P_{31}~(1750)$
comes out much below the PDG listings.\\

\noindent{\bf D-waves}\\ For both isospin-1/2 and 3/2 D-waves our resonance
parameters generally agree with the PDG values, except for the fact that we
do not find a pole corresponding to the $D_{13}~(1700)$.\\

\noindent{\bf F-waves}\\
Besides the fact that we can not describe the $F_{17}$ channel,
our results for the $F-$wave resonance parameters are in good
agreement with the PDG listings.

\section{Summary and conclusion}

In earlier work we constructed a meson-exchange model for the $\pi N$
interaction which describes the $\pi N$ elastic scattering data up to pion
laboratory energy of 400~MeV~\cite{hung94,hung01}. Our approach was based on a
three-dimensional reduction scheme of the Bethe-Salpeter equation for a model
Lagrangian involving $\pi, N, \Delta, \sigma$, and $\rho$ fields. This model
was later extended to energies up to 2~GeV in the $S_{11}$ channel by
explicitly including the $\eta N$ channel and several higher
resonances~\cite{chen03}. The influence of the $2\pi$ channels was accounted
for by adding a phenomenological term in the resonance propagator. Good
agreement was obtained with the data from the $\pi N\rightarrow\eta N$ reaction
and pion photoproduction.\\

In the present work, the hadron-exchange coupled channels model has been
further extended to energies of 2~GeV and partial wave channels including the
$F-$waves. We have assumed that all the resonances observed in $\pi N$
scattering are fundamentally three-quark states dressed by the coupling to the
meson-nucleon continuum. Based on such a scheme, we are able to achieve a very
good description of the $\pi N$ elastic scattering amplitudes in all the
partial-waves and over the energy range up to 2~GeV, except for the $F_{17}$
channel. However, the fit to the data requires four additional resonances with
very large widths, $S_{11}(1878), D_{13}(2152), P_{13}(2204)$, and
$P_{31}(2100)$, which are not listed by the PDG~\cite{PDG2006}. \\

We have developed a scheme to extract the parameters of overlapping resonances
in a completely symmetrical way with respect to the resonances. This scheme
allows us to include the dressing of each particular resonance due to all the
other resonances in the same channel. We have chosen to define the resonance
energy such that the effect of vertex dressing is not included in the
self-energy of a resonance, contrary to many previous investigations.
Furthermore, the pole positions and the residues of the scattering amplitudes
at the pole have been determined by means of the speed-plot technique. The
comparison of the extracted resonance parameters with the PDG values yields a
qualitative agreement in general but considerable discrepancies in some cases,
in particular for the widths and residues of some higher resonances. Further
investigations will be necessary to understand these differences in detail.\\

The $\pi N$ model developed in this work will be used to study the
meson cloud effects on the electromagnetic transition form factors
of the higher resonances. It will also allow us to extract the
helicity amplitudes of all resonances in a more consistent and
reliable way.

\vspace{1.0cm} \centerline{\bf Acknowledgment} S.S.K. wishes to
acknowledge the financial support from the National Science
Council of ROC for his visits to the Physics Department of
National Taiwan University. The work of S.N.Y. is supported in
part by the NSC/ROC under grant No.~NSC095-2112-M022-025. We are
also grateful for the support of the the Deutsche
Forschungsgemeinschaft through the SFB~443, by joint project
NSC/DFG 446 TAI113/10/0-3 and  by the joint Russian-German
Heisenberg-Landau program.\\

\newpage

\begin{table}[htb]
\centering \setlength{\textwidth}{50mm} \caption{Bare
($M_R^{(0)}$) and physical ($M_R$) resonance masses as well as
total widths $\Gamma_R$, all in units of MeV, single pion
branching ratios $\beta^{1\pi}_R$, and background phases $\phi_R$
of Eq.~(\ref{phi_Ri}) for isospin-1/2 resonances. Upper lines: our
results, lower lines: results of the PDG~\cite{PDG2006}.}
\vspace{0.5cm}
\begin{tabular}{|l|ccccc|}
\hline
 $N^*$ & $M_R^{(0)}$ & $M_R$ & $\Gamma_R$ & $\beta_R^{1\pi}(\%)$ &
 $\phi_R$(deg)\\
\hline
$P_{11}(1440)$  & 1612 & 1418 & 436 & 44 & 32   \\
 $****$&  &$1445\pm 25$& $325\pm 125$ & $65\pm 10$ & \\
 \hline
$D_{13}(1520)$  & 1590 & 1520 & 94  & 62 & 1.2   \\
 $****$ &  &$1520\pm 5$& $115\pm 15$ & $60\pm 5$ &   \\
 \hline
$S_{11}(1535)$  & 1559 & 1520 & 130 & 43 & 20   \\
 $****$ &  &$1535\pm 10$& $150\pm 25$ & $45\pm 10$ &   \\
 \hline
$S_{11}(1650)$  & 1727 & 1678 & 200 & 73 & 24    \\
 $****$ &  &$1655\pm 10$& $165\pm 20$ & $77\pm 17$ &  \\
\hline
$D_{15}(1675)$  & 1710 & 1670 & 154 & 18 & 49   \\
 $****$  &  &$1675\pm 5$& $147\pm 17$ & $40\pm 5$ &   \\
 \hline
$F_{15}(1680)$  & 1748 & 1687 & 156 & 67 & 7.9   \\
 $****$ &  &$1685\pm 5$& $130\pm 10$ & $67\pm 2$ &   \\
 \hline
$D_{13}(1700)$  & 1753 & 1747 & 156 & 5  & -1    \\
 $***$  &  &$1700\pm 50$& $100\pm 50$ & $10\pm 5$ &   \\
 \hline
$P_{11}(1710)$  & 1798 & 1803 & 508 & 32 & 40    \\
 $***$   &  &$1710\pm 30$& $180\pm 100$ & $15\pm 5$ &   \\
 \hline
$P_{13}(1720)$  & 1725 & 1711 & 278 & 13 & 0     \\
 $****$   &  &$1725\pm 25$& $225\pm 75$ & $15\pm 5$ &  \\
 \hline
$P_{13}(1900)$  & 1922 & 1861 & 1000& 18 &-3.5   \\
 $**$  &  &$1879\pm 17$& $498\pm 78$ & $26\pm 6$ &   \\
 \hline
$F_{15}(2000)$  & 1928 & 1926 & 58  & 4  & 18    \\
 $**$  &  &$1903\pm 87$& $490\pm 310$ & $8\pm 5$ &   \\
 \hline
$D_{13}(2080)$  & 1972 & 1946 & 494 & 15 & 5    \\
 $**$   &  &$1804\pm 55$& $450\pm 185$ & $\sim 4$ &  \\
 \hline
$S_{11}(xxx) $  & 1803 & 1878 & 508 & 41 & -5    \\
 \hline
$S_{11}(2090)$  & 2090 & 2124 & 388 & 37 & -18  \\
 $*$   &  &$2180\pm 80$& $350\pm 100$ & $18\pm 8$ &   \\
 \hline
$P_{11}(2100)$  & 2196 & 2247 & 1020& 42 & 32    \\
 $*$  &  &$2125\pm 75$& $260\pm 100$ & $12\pm 2$ &  \\
 \hline
$D_{13}(xxx) $  & 2162 & 2152 & 292 & 14 & 7    \\
\hline
$P_{13}(xxx )$  & 2220 & 2204 & 406 & 15 & -4    \\
\hline
$D_{15}(2200)$  & 2300 & 2286 & 532 & 16 & 8     \\
 $**$ &  &$2180\pm 80$& $400\pm 100$ & $10\pm 3$ &   \\
 \hline
\end{tabular}\label{table:param-1/2}
\end{table}

\begin{table}[htb]
\centering \setlength{\textwidth}{50mm} \caption{Bare
($M_R^{(0)}$) and physical ($M_R$) resonance masses as well as
total widths $\Gamma_R$, all in units of MeV, single pion
branching ratios $\beta^{1\pi}_R$, and phases $\phi_R$ for
isospin-3/2 resonances. Notation as in
Table~\ref{table:param-1/2}. }\vspace{0.5cm}

\begin{tabular}{|l|ccccc |}
\hline
 $N^*$ & $M_R^{(0)}$ & $M_R$ & $\Gamma_R$ & $\beta_R^{1\pi}(\%)$ &
 $\phi_R$(deg)\\
\hline
$P_{33}(1232)$  & 1425 & 1233 & 132 & 100 & 12   \\
 $****$&  &$1232\pm 1$& $118\pm 2$ & $100$ &   \\
 \hline
$P_{33}(1600)$  & 1575 & 1562 & 216  & 6 & -9   \\
 $***$ &  &$1600\pm 100$& $350\pm 100$ & $17\pm 7$ &   \\
 \hline
$S_{31}(1620)$  & 1654 & 1616 & 160 & 32 & -41   \\
 $****$ &  &$1630\pm 30$& $142\pm 18$ & $25\pm 5$ &   \\
 \hline
$D_{33}(1700)$  & 1690 & 1650 & 260 & 15 & -5   \\
 $****$ &  &$1710\pm 40$& $300\pm 100$ & $15\pm 5$ &   \\
\hline
$P_{31}(1750)$  & 1765 & 1746 & 554 & 4 & -24   \\
 $*$  &  &$1744\pm 36$& $300\pm 120$ & $8\pm 3$ &   \\
 \hline
$S_{31}(1900)$  & 1796 & 1770 & 430 & 8 & -44    \\
 $**$ &  &$1900\pm 50$& $190\pm 50$ & $2\pm 1$ &  \\
 \hline
$F_{35}(1905)$  & 1891 & 1854 & 534 & 11  & -12    \\
 $****$  &  &$1890\pm 25$& $335\pm 65$ & $12\pm 3$ &  \\
 \hline
$P_{31}(1910)$  & 1953 & 1937 & 226 & 14 & -21   \\
 $****$   &  &$1895\pm 25$& $230\pm 40$ & $22\pm 7$ &   \\
 \hline
$P_{33}(1920)$  & 1856 & 1827 & 834 & 12 & 3     \\
 $***$   &  &$1935\pm 35$& $220\pm 70$ & $12\pm 7$ &   \\
 \hline
$D_{35}(1930)$  & 2100 & 2068 & 426& 15 &-20    \\
 $***$  &  &$1960\pm 60$& $360\pm 140$ & $10\pm 5$ &   \\
 \hline
$D_{33}(1940)$  & 2100 & 2092 & 310  & 6  & -10   \\
 $*$  &  &$2057\pm 110$& $460\pm 320$ & $18\pm 12$ &  \\
 \hline
$F_{37}(1950)$  & 1974 & 1916 & 338 & 47 &  13     \\
 $****$   &  &$1932\pm 17$& $285\pm 50$ & $40\pm 5$ &   \\
 \hline
$F_{35}(2000) $  & 2277 & 2260 & 356 & 11 & -26    \\
 $**$   &  &$2200\pm 125$& $400\pm 125$ & $16\pm 5$ &   \\
 \hline
$P_{31}(xxx)$  & 2160 & 2100 & 492 & 35 & -25   \\
 \hline
$S_{31}(2150)$  & 2118 & 1942 & 416 & 70 & -44   \\
 $*$  &  &$2150\pm 100$& $200\pm 100$ & $8\pm 2$ &   \\
\hline
\end{tabular}\label{table:param-3/2}
\end{table}

\begin{table}[htb]
\centering \setlength{\textwidth}{50mm} \caption{Pole positions
$W_p-\frac{1}{2}i \Gamma_p$ and absolute values of the residues $\mid r \mid$
at the pole, all in MeV, as well as the phases $\theta$ of the residues for
isospin-1/2 resonances. Notation as in
Table~\ref{table:param-1/2}.}\vspace{0.5cm}

\begin{tabular}{|l|cccc |}
\hline
 $N^*$ & $W_p$ & $\Gamma_p$ & $\mid r \mid$ & $\theta$(deg)  \\
\hline
$P_{11}(1440)$  & 1366 & 179 & 47 & -87   \\
 $****$ &$1365\pm 15$& $190\pm 30$ & $46\pm 10$ & $-100\pm 35$  \\
 \hline
$D_{13}(1520)$  & 1516 & 123 & 40 & -6     \\
 $****$ & $1510\pm 5$& $114\pm 10$ & $35\pm 3$ &$-10\pm 4$   \\
 \hline
$S_{11}(1535)$  & 1449 &  67 & 11 & -46    \\
 $****$ & $1510\pm 20$& $170\pm 80$ & $96\pm 63$ & $ 15\pm 45$   \\
 \hline
$S_{11}(1650)$  & 1642 & 97 & 21 & -73    \\
 $****$ & $1655\pm 15$& $165\pm 15$ & $55\pm 15$ & $-75\pm 25$   \\
\hline
$D_{15}(1675)$  & 1657 & 132 & 24 & -22    \\
 $****$ & $1660\pm 5$& $137\pm 12$ & $29\pm 6$ & $-30\pm 10$   \\
 \hline
$F_{15}(1680)$  & 1663 & 115 & 38 & -28    \\
 $****$ & $1672\pm 8$& $122\pm 12$ & $38\pm 6$ & $-23\pm 7 $   \\
 \hline
$D_{13}(1700)$  & not seen & not seen & not seen & not seen     \\
 $***$  & $1680\pm 50$& $100\pm 50$ & $6\pm 3$ & $0\pm 50$   \\
 \hline
$P_{11}(1710)$  & 1721 & 185 & 5 & -163    \\
 $***$   & $1720\pm 50$& $230\pm 150$ & $10\pm 4$ & $-175\pm 35$   \\
 \hline
$P_{13}(1720)$  & 1683 & 239 & 15 & -64     \\
 $****$   &  $1675\pm 15$& $195\pm 80$ & $13\pm 7$ & $-139\pm 51$   \\
 \hline
$P_{13}(1900)$  & 1846 & 180 & 7 & -75    \\
 $**$  & not listed & not listed & not listed & not listed   \\
 \hline
$F_{15}(2000)$  & 1931 & 62 & 1.3  & -272     \\
 $**$  & not listed & not listed & not listed & not listed   \\
 \hline
$D_{13}(2080)$  & 1834 & 210 & 13 & -134    \\
 $**$  & $1950\pm 170$& $200\pm 80$ & $27\pm 22$&$\sim 0$   \\
 \hline
$S_{11}(2090)$  & 2065 & 223 & 16 & -138    \\
 $*$   & $2150\pm 70$& $350\pm 100$ & $40\pm 20$ & $0\pm 90$  \\
 \hline
$P_{11}(2100)$  & 1869 & 238 & 7 & -216     \\
 $*$  & $2120\pm 240$& $240\pm 80$ & $14\pm 7$ & $35\pm 35$  \\
 \hline
$D_{15}(2200)$  & 2188 & 238 & 21 & -27    \\
 $**$ & $2100\pm 60$& $360\pm 80$ & $20\pm 10$ & $-90\pm 50$   \\
 \hline
\end{tabular}\label{table:pole-1/2}
\end{table}

\begin{table}[htb]
\centering \setlength{\textwidth}{50mm} \caption{Pole positions
$W_p-\frac{1}{2}i \Gamma_p$ and absolute values of the residues $\mid r \mid$
at the pole, all in MeV, as well as the phases of the residues $\theta$ for
isospin-3/2 resonances. Notation  as in
Table~\ref{table:param-1/2}.}\vspace{0.5cm}

\begin{tabular}{|l|cccc|}
\hline
 $N^*$ & $W_p$ & $\Gamma_p$ & $\mid r \mid$ & $\theta$(deg) \\
\hline
$P_{33}(1232)$  & 1218 & 89 & 42 & -35    \\
 $****$ &$1210\pm 1$& $100\pm 2$ & $53\pm 2$ & $-47\pm 1$   \\
 \hline
$P_{33}(1600)$  & 1509 & 236 & 35 & -197    \\
 $***$ & $1600\pm 100$& $300\pm 100$ & $17\pm 4$ &$-150\pm 30$   \\
 \hline
$S_{31}(1620)$  & 1598 &  136 & 22 & -99    \\
 $****$ & $1600\pm 10$& $118\pm 3$ & $16\pm 3$ & $ -110\pm 20$   \\
 \hline
$D_{33}(1700)$  & 1609 & 133 & 9.5 & -52   \\
 $****$ & $1650\pm 30$& $200\pm 40$ & $13\pm 3$ & $-20\pm 25$   \\
\hline
$P_{31}(1750)$  & 1729 & 70  & 1 & -123    \\
 $*$            & 1748 & 524 & 48  &  158   \\
 \hline
$S_{31}(1900)$  & 1775 & 36  & 1 & -166     \\
 $**$ & $1870\pm 40$& $180\pm 50$ & $10\pm 3$ & $-20\pm 40 $   \\
 \hline
$F_{35}(1905)$  & 1771 & 190 &  11 & -47    \\
 $****$  & $1830\pm 5$& $280\pm 20$ & $25\pm 8$ & $-50\pm 20$   \\
 \hline
$P_{31}(1910)$  & 1896 & 130 & 6 & -118     \\
 $****$   & $1880\pm 30$& $200\pm 40$ & $20\pm 4$ & $-90\pm 30$  \\
 \hline
$P_{33}(1920)$  & 2149 & 400 & 38 & -59   \\
 $***$   &  $1900\pm 50$& $300\pm 100$ & $24\pm 4$ & $-150\pm 30$   \\
 \hline
$D_{35}(1930)$  & 1992 & 270 & 18 & -75     \\
 $***$   &  $1900\pm 50$& $265\pm 95$ & $18\pm 6$ & $-20\pm 40$   \\
 \hline
$D_{33}(1940)$  & 2070 & 267 & 7  & -31    \\
 $*$   &  $1900\pm 100$& $200\pm 60$ & $24\pm 4$ & $135\pm 45$  \\
 \hline
$F_{37}(1950)$  & 1860 & 201 & 43 & -45   \\
 $****$  & $1880\pm 10$& $240\pm 20$ & $50\pm 7$&$ -33\pm 8$   \\
 \hline
$F_{35}(2000)$  & 2218 & 219 & 11 & -36    \\
 $**$   & $2150\pm 100$& $350\pm 100$ & $16\pm 5$ & $150\pm 90$ \\
 \hline
$S_{31}(2150)$  & 2012 & 148 & 6 & -155     \\
 $*$  & $2140\pm 80$& $200\pm 80$ & $7\pm 2$ & $-60\pm 90$ \\
 \hline
 \end{tabular}\label{table:pole-3/2}
\end{table}

\end{document}